\documentclass{aastex631}
\usepackage{amsmath}

\begin{document}

%\title{Decoding the Cosmic Orchestra: Deep Learning Analysis of Precessing Binary Black Holes in Noisy LIGO Data }
\title{Reconstruction of binary black hole harmonics in LIGO using deep learning} 

\author[0000-0001-8700-3455]{Chayan Chatterjee}
\affiliation{Department of Physics and Astronomy, Vanderbilt University\\ 2201 West End Avenue, Nashville, Tennessee - 37235,}
\affiliation{Data Science Institute, Vanderbilt University\\ 1400 18th Avenue South Building, Suite 2000, Nashville, Tennessee - 37212, }

\author[0000-0003-1007-8912]{Karan Jani}
\affiliation{Department of Physics and Astronomy, Vanderbilt University\\ 2201 West End Avenue, Nashville, Tennessee - 37235,} 
%\kpj{add DSI}



\begin{abstract}

%Gravitational wave signals detected by the LIGO and Virgo interferometers are dominated by noise, primarily originating from instrumental and terrestrial sources. This noise, characterized by its non-Gaussian and non-stationary nature, is further complicated by the presence of sharp transients, known as glitches, which often mimic the morphology of gravitational wave chirps. In this work, we employ deep learning to tackle the challenge of reconstructing true astrophysical gravitational wave signals from noisy strain data. Our model is trained and tested on fully precessing binary black hole waveforms, inclusive of higher-order modes, and injected into real detector noise from LIGO's third observation run. Furthermore, we test our model on real gravitational wave events detected during the first three observation runs and achieve a high degree of fidelity between our model's predictions and the maximum likelihood reconstructions obtained by coherent Wave Burst and LALInference analyses.

Gravitational wave signals from coalescing compact binaries in the LIGO and Virgo interferometers are primarily detected by the template-based matched filtering method. While this method is optimal for stationary and Gaussian data scenarios, its sensitivity is often affected by non-stationary noise transients in the detectors. Moreover, most of the current searches do not account for the effects of precession of black hole spins and higher-order waveform harmonics, focusing solely on the leading-order quadrupolar modes. This limitation impacts our search for interesting astrophysical sources, such as intermediate-mass black hole binaries and hierarchical mergers. Here we show for the first time that deep learning can be used for accurate waveform reconstruction of precessing binary black hole signals with higher-order modes. This approach can also be adapted into a rapid trigger generation algorithm to enhance online searches. Our model, tested on simulated injections in real LIGO noise from the third observing run (2019-2020) achieved high-degree of overlap with injected signals. This accuracy was consistent across a wide range of black hole masses and spin configurations chosen for this study. When applied to real gravitational wave events, our model’s reconstructions achieved between 85\% and 98\% overlap with those obtained by Coherent WaveBurst (unmodeled) and LALInference (modeled) analyses. These results suggest that deep learning is a potent tool for analyzing signals from a diverse catalog of compact binaries.

%\kpj{Gravitational wave signals from black holes and neutron stars in LIGO-Virgo detectors are dominated by non-Gaussian and non-stationary instrumental and environmental noise. In this work, we demonstrate for the first time a direct deep learning application to the noisy LIGO-Virgo data for the state-of-the-art gravitational wave models that include higher harmonics, spin precession and long-duration signals.   Karan is still wriring...}

\end{abstract}



\section{Introduction} \label{sec:intro}

In the last three observation runs of LIGO (\cite{LIGO}) and Virgo (\cite{Virgo}), the number of gravitational wave (GW) detections from coalescing binaries has increased to almost a hundred. These discoveries have ushered in a new era in astrophysics, providing a novel means to probe the universe's most energetic processes. The recent GW transient catalogue, GWTC-3 (\cite{GWTC-3}), marks a significant milestone in the ongoing efforts of the LIGO-Virgo-KAGRA (LVK) Collaboration (\cite{LIGO,Virgo,KAGRA,KAGRA_1}), contributing to the increasing rate of GW detections in each observation run. With enhanced sensitivity in the interferometers, the detection rate during the ongoing fourth observation run (O4) has risen to about one every few days, further enriching the rapidly growing catalogue of compact object mergers (\cite{GraceDB}). \\

In GW astronomy, the classic approach to confirming the presence of signals in noisy strain timeseries is the application of a matched filtering technique (\cite{Matched_filtering,MF1,MF2,MF3}) which is optimal in scenarios where the background noise is Gaussian and stationary. This technique, however, requires substantial computational resources to generate millions of filter templates based on varying source masses ($m_{i}$) and spins ($\vec{\chi}_{i}$), to accurately identify real signals. However, actual interferometer data is not only non-Gaussian and non-stationary, but is also frequently corrupted by sharp noise features called `glitches' which can mimic true GW transients. Additionally, GW templates currently used in most searches for binary black holes (BBHs) do not include the effects of precession of the orbital plane and higher-order modes. Recent work by \cite{Detection_HM_2}, based on the previous study by \cite{Detection_HM_1} represents the first major step in this direction. Studies have shown that this exclusion can lead to considerable fractional losses in sensitive volume for binaries with mass ratios, m$_{1}$/m$_{2}$ $\geq$ 4 and total masses, $M=m_{1}+m_{2}$, exceeding 100 M$_{\odot}$ in the detector frame (\cite{EffectofHM1,EffectofHM2,EffectofHM3,EffectofHM4}). In fact, for systems with the most extreme cases of precession, this loss in sensitivity persists even for sources which are face-on to the detector (\cite{EffectofHM5}). \cite{Bustillo} reports that ignoring these aspects can result in a loss of sensitive volume of approximately 15\% - 25\%. Therefore, with the evolving landscape of GW astronomy, there is a growing need to develop faster and reliable techniques to complement existing data analysis and search methods. \\

Deep learning methods are particularly well-suited for GW data analysis due to its ability to learn complex patterns and features from large datasets, effectively handling the non-stationary noise characteristics of raw detector strains. Several applications of deep learning in GW data analysis have been reported, including the use of convolutional neural networks (CNNs) (\cite{CNN_1,CNN_2}) for rapid signal detection (\cite{Gabbard_detection, detection2, detection7,Beveridgedetection}) and glitch classification (\cite{GravitySpy,glitch1,glitch2}), normalizing flows (\cite{NormalizingFlow}) and conditional variational autoencoders (\cite{CVAE}) for rapid parameter estimation (\cite{DINGO_IS, Gabbard, Chua, GW-SkyLocator}) and total variation methods (\cite{denoising3}), dictionary learning (\cite{TorresForne}), denoising autoencoders (\cite{Bacon_denoising, DanielGeorge_denoising, Chatterjee_2021}) and \texttt{WaveNet}-based architectures (\cite{WaveNet} \cite{Wei&Huerta_denoising}) for GW signal extraction from noise. \\

In this article, we present a study on GW signal reconstruction from noise using an enhanced version of the model originally introduced in \cite{Chatterjee_2021}. This iteration of the model uniquely incorporates both precession and higher-order modes, previously unaddressed aspects, from fully precessing black hole binaries. Our analysis is done on simulated injections in real LIGO noise recorded during the third observation run (O3), as well as on actual GW event data. The broad aim of this research is to boost GW searches by serving as a trigger generation algorithm for potential astrophysical candidates. This paper is organized as follows: Section 1.1 discusses the astrophysical implications of detecting GW higher-order modes and signatures of precession in binary orbits. Section 2 describes the modifications made to our model from \cite{Chatterjee_2021}, as well as the process of generating training and test datasets. Section 3 presents the results obtained using this updated model on simulated injections in noise from the third observation run (O3) and on real GW events detected during the last three observation runs. Section 4 outlines our conclusions and the potential for future studies.

\subsection{Higher-order modes and precession}

%There are 15 parameters which determine the evolution of GW signals from BBHs in quasicircular orbits. Some of these are intrinsic parameters, like the individual masses $m_{i}$ and dimensionless spins $\vec{\chi}_{i}$ of the component black holes. The other parameters, called extrinsic parameters define the position and orientation of the binary system with respect to the detector. Following the notation in \cite{Bustillo}, the GW strain can be expressed in terms of the intrinsic and extrinsic parameters as follows:

%\begin{equation}
%    h=\frac{F}{d_{L}}(\cos \psi \Re(\mathcal{H})+\sin \psi \Im(\mathcal{H})),
%\label{eq:1}
%\end{equation}

%where $\mathcal{H}$ represents the complex GW strain $h_{+}-i h_{\times}$. $\Re(\mathcal{H})$ and $\Im(\mathcal{H})$ denotes the real and imaginary parts of the complex strain. $\psi $ is an effective polarization angle term defined as $\tan \psi=F_{\times} / F_{+}$, where $F_{+} $ and $F_{\times} $ are the antenna beam pattern functions of the detectors, which can be used to obtain the global factor $F$. 

GW signals can be decomposed into a sum of modes $h_{\ell, m}$ weighted by spin -2 spherical harmonics $Y_{\ell, m}$ as follows (\cite{Bustillo}):
\begin{equation}
\mathcal{H} =\sum_{\ell \geq 2} \sum_{m=-\ell}^{m=\ell} Y_{\ell, m}^{-2}(\vartheta, \varphi) h_{\ell, m}\left(\Xi ; t-t_{c}\right).
\end{equation}
Here, $\vartheta$ and $\varphi$ are polar and azimuthal angles which define the location of the detector in a reference frame fixed at the center of mass of the binary. The parameter $\boldsymbol{\Xi}$ collectively denotes the intrinsic parameters of the binary, i.e., $\boldsymbol{\Xi}=\left(m_{1}, m_{2}, \vec{\chi}_{1}, \vec{\chi}_{2}\right)$, and $t_{c}$ denotes the coalescence time. The GW strain is dominated by the $(\ell, m)=(2, \pm 2)$ modes for non-precessing binaries, while the higher-order modes only make significant contributions to the signal at the merger and ringdown stages (\cite{EffectofHM3, EffectofHM4, EffectofHM1, HM_ringdown, EffectofHM2}).\\

%The total angular momentum $\vec{J}$ of the binary can be expressed as a sum of the orbital angular momentum $\vec{L}$ and the component spins, $\chi_{1} $ and $\chi_{2} $ as,

%\begin{equation}
%    \vec{J}=\vec{L}+\vec{\chi}_{1} m_{1}^{2}+\vec{\chi}_{2} m_{2}^{2} \equiv \vec{L}+\vec{S},
%\end{equation}

Precession occurs when one or both black hole spins are misaligned with respect to the orbital angular momentum of the binary. This causes the orbital plane and the spins to precess around the direction of the total angular momentum. The effective inspiral spin, $\chi_{\text {eff }}$, and the effective spin precession, $\chi_{p}$, are used to encode information about the in-plane and parallel components of the black hole spins. These parameters are defined as follows (\cite{Precession1,Precession2}):

\begin{equation}
    \chi_{p}=\max \left(\chi_{1}\text{sin}\theta_{1}, \left(\frac{4q + 3}{4 + 3q}\right) q\chi_{2}\text{sin}\theta_{2}\right), \quad \chi_{\mathrm{eff}}=\frac{\chi_{1}\text{cos}\theta_{1} +\chi_{2}q\text{cos}\theta_{2}}{1 + q},
\end{equation}

%\begin{equation}
%    \chi_{p}=\max \left(\chi_{1}^{\perp}, \frac{A}{q^{2}} \chi_{2}^{\perp}\right), \quad \chi_{\mathrm{eff}}=\frac{\chi_{1}^{\|} m_{1}+\chi_{2}^{\|} m_{2}}{m_{1}+m_{2}},
%\end{equation}

%where $q = m_{2}/m_{1}, m_{2} \leq m_{1}$, $A=2+1.5 q$, $\vec{\chi}_{i}=\vec{\chi}_{i}^{\|}+\vec{\chi}_{i}^{\perp}$, with $\vec{\chi}_{i}^{\|} $ and $\vec{\chi}_{i}^{\perp} $ denoting the parallel and perpendicular components of $\vec{\chi}_{i} $. The effect of precession of the orbital plane, characterized by a non-zero $\chi_{p} $ is to induce amplitude and phase modulations of the $h_{\ell, m}$ modes. \\

where $q = m_{2}/m_{1}, m_{2} \leq m_{1}$, $\chi_{1} $ and $\chi_{2} $ are the primary and secondary spin magnitudes,  $\theta_{1} $ and $\theta_{2}$ are the primary and secondary tilt angles respectively. \\

Detecting higher-order harmonics in GW signals can significantly enhance our understanding of binary system characteristics. This is due to the varied dependencies of different modes on the binary's orientation, which help resolve degeneracies between specific source parameters. Moreover, observing higher-order modes from the ringdown stage provides an opportunity to rigorously test the predictions of General Relativity (\cite{HM_ringdown}). This is particularly relevant for massive binaries involving intermediate mass black holes where the merger frequency may align with the peak sensitivity of current detectors (\cite{IMBH_HM, LISA_IMBH_HM}). Additionally, these modes offer insights into critical astrophysical phenomena such as recoil kicks, which play a key role in black hole formation and evolution (\cite{BH_kicks_1, BH_kicks_2}). In precessing systems, these kick velocities can reach up to 5000 km/s, surpassing the escape velocities of even the largest galaxies. Understanding this phenomenon is vital for predicting event rates for future missions like the Laser Interferometer Space Antenna (LISA) (\cite{LISA_1, LISA_2}). 

\section{Method}

In \cite{Chatterjee_2021}, a CNN and a Long Short Term Memory (LSTM) network (\cite{LSTM}) were integrated in an encoder-decoder architecture to reconstruct pure GW signals from noisy strain data. The model's input was a 0.25-second long noisy strain time series, from which the network generated either `noise-less' pure templates (for astrophysical signals) or a vector of random fluctuations around zero amplitude (for pure noise). In the present work, we have expanded upon the model to reconstruct 1-second long BBH signals, incorporating higher-order modes and precession. In this section, we describe the modifications to the model architecture in \cite{Chatterjee_2021} implemented for this work and the data generation method used to simulate our training and test sets.

\subsection{Training and testing data}

The `clean' GW signals used for this study are simulated BBH waveforms of masses between 5 and 100 M$_{\odot}$ in the detector frame. The spin magnitudes of the component black holes are uniformly sampled between 0 and 0.99 in the $x $, $ y$ and $z$ directions in a Cartesian co-ordinate system. For the tilt angles, a uniform solid angle prior was chosen which results in uniform polar and azimuthal angle priors for the component black holes between the ranges (0, $\pi $) and (0, $2\pi $) respectively. The IMRPhenomXPHM waveform approximant (\cite{IMRPhenomXPHM}) was used for the simulations, which incorporates the $(l,m)$ = (2,1), (3,2), (3,3) and (4,4) spherical harmonic modes, besides the dominant (2,2) mode. The sampling of additional parameters, such as sky coordinates, coalescence phase, polarization angle, and cosine of inclination angle, is done uniformly across their full ranges. The strain signals were generated with optimal matched filtering signal-to-noise ratio (SNR), as defined in Eq.~2.3 in \cite{SNR}, uniformly distributed between 8 and 30. The lower and upper frequency cutoffs chosen for the waveform simulations are 20 $\mathrm{~Hz}$  and 2048 $\mathrm{~Hz}$ respectively. \\

The noisy strain data is generated by injecting the GW waveforms into different segments of LIGO Livingston noise recorded in the second half of the third observing run (O3b). We generated 1 second-long strains, an upgrade from the 0.25-second duration signals used in \cite{Chatterjee_2021}. Care was taken to choose segments that are free from hardware signal injections used for testing and calibrating the detector. The noise data was downsampled from the original sampling rate of 16384 Hz to 2048 Hz, which aligns with a Nyquist frequency of 1024 Hz. The GW waveforms are then injected into the noise segments in a randomized manner, employing random time shifts uniformly distributed between (-0.1, 0.1) seconds from an arbitrarily chosen `merger time' of 0.8 seconds from the start of the 1-second signal. The noise data is sourced from the publicly available datasets by the LVK Collaboration through the Gravitational Wave Open Science Center (GWOSC) (\cite{GWOSC}). \\

The input data is preconditioned into overlapping subsequences in a similar manner as in \cite{Chatterjee_2021}. However, we have increased from 4 to 15 overlapping timesteps. With a sampling rate of 2048 Hz for 1 second long signals, this results in 2063 overlapping subsequences of 15 timesteps each. We pass each subsequence through one copy of the encoder network as before and then concatenate the feature vectors generated from each copy of the encoder to be processed by the bidirectional LSTM decoder layers which outputs the reconstructed waveform. The labels presented to the model during training are the `clean' GW waveforms which are padded appropriately to ensure that they have the same shape as the output of the decoder. We also whiten the strain data and the template waveforms and normalize the amplitudes between -1 and 1. Whitening was done to adjust for the difference in the strength of the signals at different frequencies by dividing the Fourier transform of the signals by an estimate of the amplitude spectral density of the noise. During whitening, we also applied a Tukey window with $\alpha=1 / 8$. \\

\subsection{Model architecture and training}

The model architecture in \cite{Chatterjee_2021} was a CNN-LSTM denoising autoencoder, consisting of a CNN encoder and a LSTM decoder (Fig.~1 in \cite{Chatterjee_2021}). Denoising autoencoders are a type of neural network used for the task of removing noise from corrupted input data. They are trained to reconstruct a clean or uncorrupted version of the input from noisy signals, learning to capture the important features while discarding the noise. This makes them particularly useful in applications like image and audio processing, where they can enhance the data quality by filtering out unwanted disturbances. When applied to GWs, denoising autoencoders can accurately extract pure GW waveforms from noisy strain data, provided they are trained with a sufficient number of strain and signal pairs, serving as the input data and labels respectively. \\

\begin{figure}
\gridline{\fig{AWaRE.png}{0.55  \textwidth}{(a)}
          \fig{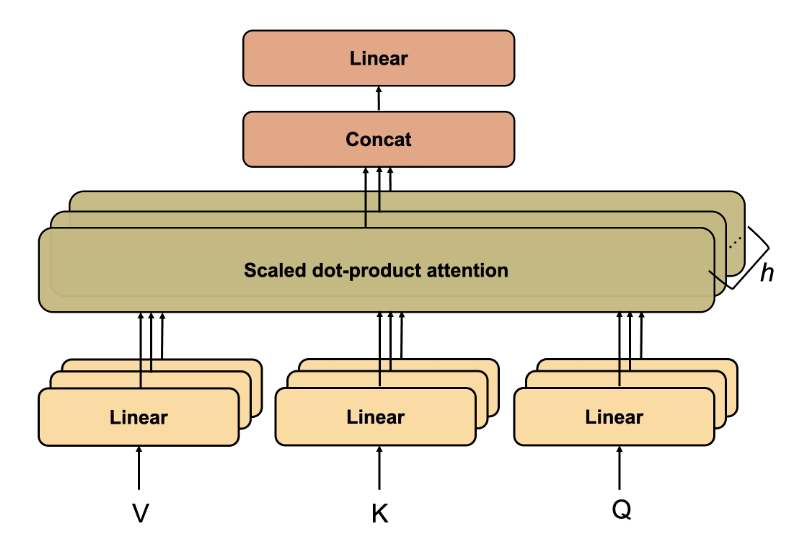}{0.40\textwidth}{(b)}}
\caption{\label{fig:AWaRe} (a) Architecture of \texttt{AWaRe} model. The input is a noisy strain GW sugnal which is reshaped into overlapping subsequences of 15 timesteps each that passes through a copy of the encoder network consisting of 1-D convolutional (Conv1D) and MHA layers. The decoder consists of LSTM layers that reconstruct the pure waveform. (b) Components of the MHA module which implements the scaled dot-product on Q, K and V matrices and concatenates the output.}
\end{figure}

In the updated version of the model (shown in Fig.~\ref{fig:AWaRe}), which we have named \textbf{\texttt{AWaRe}} -- \textbf{A}ttention-boosted \textbf{Wa}veform \textbf{Re}construction network, we have incorporated a multi-head attention module, first introduced by \cite{Attention}, as an important component of the Transformer architecture. Here we implement multi-head attention at the end of the convolutional layers in our encoder network for improved feature extraction from noisy strain data. Multihead attention is adept at capturing long-range dependencies within the input data and allows the network to focus on different parts of the input sequence simultaneously. At each timestep of the input sequence, the attention module determines which other parts of the sequence are important and must be `paid attention to' in order to make accurate predictions. To do this, it assigns a score to the \( d \)-dimensional encoded representations of the input sequences, using a scaled dot product operation as follows:

\begin{equation}
    \text{Attention}(Q, K, V) = \text{softmax}\left(\frac{QK^T}{\sqrt{d}}\right)V,
\end{equation}

Here the Query \( Q \), Key \( K \) and Value \( V \) are matrices of the encoded representations of the input sequences having dimension \(\mathbb{R}^{l \times d}\), with \(l\) being the length of the input sequence. In our case, \(l\) = 15 since we prepare the data as overlapping sequences of 15 timesteps each, and we choose $d$ to be 32. Softmax is a function that normalizes inputs in the range (0, 1), representing weights that sum up to 1. For each element \( z_i \) in the vector, the softmax is calculated as:

\begin{equation}
    \text{softmax}(z_i) = \frac{e^{z_i}}{\sum_{j=1}^{K} e^{z_j}},
\end{equation}

Here, \( z_i \) represents the \( i^{th} \) element of the input vector \( z \), and \( K \) is the number of classes or elements in the vector. The denominator, \( \sum_{j=1}^{K} e^{z_j} \), is the sum of the exponentiated elements of the entire input vector, ensuring that the output values sum to 1 and each output value lies in the range (0, 1). The attention mechanism can be therefore be thought of as a means of weighting the input sequences with scores based on their importance. \\

The multi-head attention approach enhances this function by aggregating information from various linear projections of the original Queries, Keys, and Values. For \(H\) attention heads, the output from each head \(i\), represented as \(h_i\), is computed as:

\begin{equation}
  \text{head}_i = \text{Attention}(Q_i, K_i, V_i) = \text{softmax}\left(\frac{Q_iK_i^T}{\sqrt{d}}\right)V_i  
\end{equation}

Here, \(Q_i = QW_i^Q\), \(K_i = KW_i^K\), and \(V_i = VW_i^V\) are the projected Queries, Keys, and Values for head \(i\), ranging from 1 to \(H\), with respective learned parameters \(W_i^Q\), \(W_i^K\), \(W_i^V\). With a collection of all parameters \(W^H\), the Multi-Head Attention (MHA) of these \(H\) heads is expressed as:

\begin{equation}
    \text{MHA}(Q, K, V) = \text{Concat}(\text{head}_1, \ldots, \text{head}_H)W^H.
\end{equation}

This multi-head attention mechanism is highly efficient as it allows parallel computation for each head. Furthermore, it directly connects information from different projection subspaces, facilitating the learning of long-term dependencies in the input sequence. In addition to the MHA module, we have added extra convolutional and LSTM layers compared to the model in \cite{Chatterjee_2021}. We have also incorporated a technique known as dropout regularization, setting its probability at 0.22, to enhance the model's ability to generalize and prevent overfitting. Overfitting is a common issue in machine learning where a model, after learning the training data perfectly fails to perform well on new, unseen data. Dropout regularization works by randomly `dropping out' a subset of the model's features or neurons during the training process, at a rate of 22\% in our case. This means that during each training step, each feature or neuron has a 22\% chance of being temporarily excluded from the model. This strategy forces the model to not rely too heavily on any single feature or neuron, promoting a more robust and generalized learning. In addition, we have implemented `BatchNormalization' layers in \texttt{AWaRe} to enhance its learning efficiency and stability. Batch Normalization is a strategy to standardize the inputs to a network, ensuring they have a mean of zero and a variance of one. The primary purpose of this layer is to address the issue known as ``internal covariate shift," where the distribution of each layer's inputs changes as the parameters of the previous layers change during training. This shift can slow down the training process and make it harder for the model to converge to a solution.\\

We performed ablation studies using different network architectures and obtained the best performance metrics using the parameters listed in Table \ref{tab:model_architecture}, The activation functions used and the output shapes of the signal as it passes through the layers of the network are shown. The activation functions can be thought of as gates that decide how much of the input signal should be passed forward through the network. Essentially, these functions help the network learn complex patterns in the data. ReLU, which stands for Rectified Linear Unit, is one of the most commonly used activation functions which outputs the input directly if it is positive; otherwise, it outputs zero. \\

\begin{table}[h]
\centering
\caption{Architecture of the \texttt{AWaRe} model. The first colum lists the type of layer. Here `Time-distributed' means the same copy of the layer is applied for each of the 2063 overlapping subsequences of the signal. The second column lists the activation function used for the layer and the last column shows the shape of the signal after passing through the layer.}
\begin{tabular}{lll}
\hline
\textbf{Layer} & \textbf{Activation} & \textbf{Output Shape} \\ 
\hline
Batch Normalization & ReLU & (2063, 15, 1) \\
Conv1D (Time-distributed) - 64 units & ReLU & (2063, 15, 64)\\
Conv1D (Time-distributed) - 32 units & ReLU & (2063, 15, 32)\\
Conv1D (Time-distributed) - 32 units & ReLU & (2063, 15, 32) \\
Batch Normalization & - & (2063, 15, 32)\\
Dropout - 22\% & - & (2063, 15, 32)\\
MHA (Time-distributed) - 2 heads, 32 dims  & Softmax & (2063, 15, 32)\\
Flatten (Time-distributed) & - &(2063, 480)\\
Bidirectional LSTM - 32 units &  Tanh & (2063, 64)\\
Bidirectional LSTM - 32 units & Tanh & (2063, 64) \\
Batch Normalization & - & (2063, 64)\\
Dropout - 22\%  & - & (2063, 64)\\
Bidirectional LSTM - 32 units & Tanh & (2063, 64)\\
Bidirectional LSTM - 32 units & Tanh & (2063, 64) \\
Dense (Time-distributed) - 1 unit & Linear & (2063, 1)\\
\hline
\end{tabular}
\label{tab:model_architecture}
\end{table}

The loss function chosen for this work is the standard mean-squared error between the reconstructed waveforms and the pure signals. We found that for this work, including the Fractal Tanimoto coefficient term (\cite{Fractal_Tanimoto}) in the loss, along with mean squared error, did not result in any significant improvement of performance, as was observed in \cite{Chatterjee_2021}. We trained \texttt{AWaRe} with 50000 injection samples, 15\% of which was used as a validation set for monitoring the performance of the model during training, and an additional 1000 injection samples were generated for testing. Besides injection samples, we generated 100 pure noise samples, containing no GW injections, which were added to the training and test sets respectively. The purpose of including the pure noise samples is to train the model to distinguish real astrophysical signals from noise. The labels used for pure noise samples is simply a vector of zero amplitude. In the next section, we discuss in detail the performance of the model on injections and pure noise and the metrics used to quantify the matches between predictions and ground truths. We train \texttt{AWaRe} for 500 epochs with an initial learning rate of $10^{-3}$ which was scaled by a factor of 0.9 if the validation loss did not reduce after 25 epochs. The Adam optimizer (\cite{adam}) was used for training the model. We used a single NVIDIA DGX A100 GPU for training. The code was written in Python using the TensorFlow library (\cite{TensorFlow}).

\section{Results}

We tested \texttt{AWaRe} on injections in O3b noise and on real GW signals detected from the O1-O3 observing runs. To gauge the accuracy of our model's predictions, we compute the overlap ($\mathcal{O}$), which is traditionally used to quantify the match between different waveform models (\cite{SNR}). It is defined as,

\begin{equation}
    \mathcal{O}(h, s)=(\hat{h} \mid \hat{s}).
\end{equation}

with $\hat{h}=h(h \mid h)^{-1 / 2}$ and $\hat{s}=s(s \mid s)^{-1 / 2}$. \\

Here $h$ may refer to the reconstructed waveform produced by \texttt{AWaRe}, and $s$ either the pure template (for injections) or the maximum likelihood reconstruction obtained by coherent Wave Burst (cWB) and LALInference (for real events). \\

\subsection{Example of waveform reconstruction}
\begin{figure}
\gridline{\fig{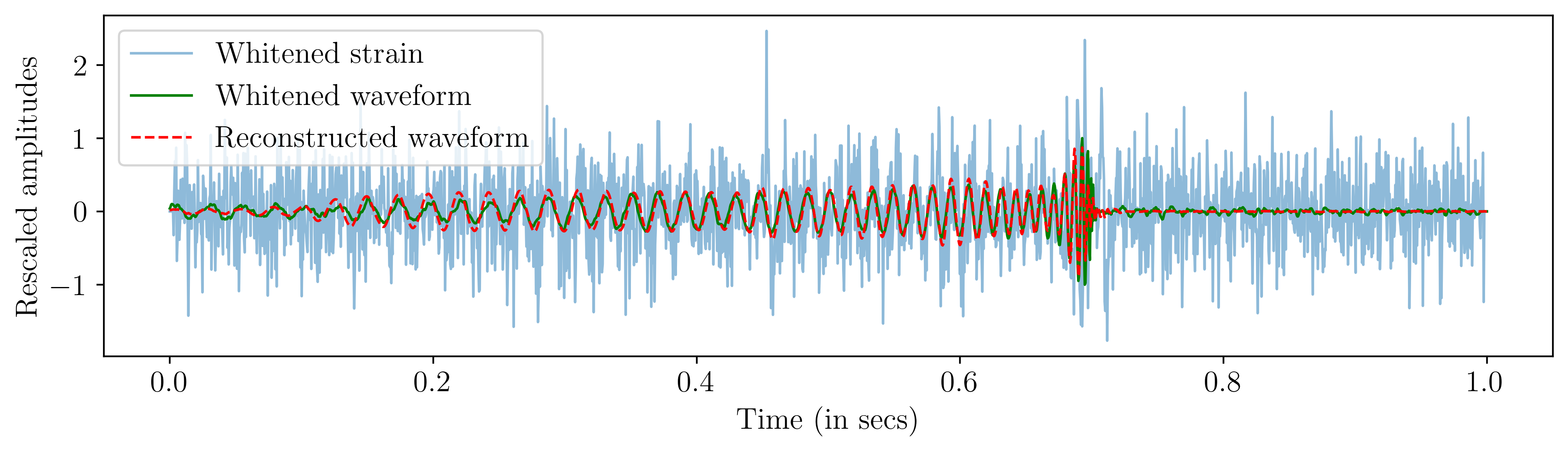}{0.8\textwidth}{(a)}}
%\gridline{\fig{Weighted_loss_Plot_m1-15_m2-14_snr-15_unlabelled.png}{0.8\textwidth}{(b)}}
\gridline{\fig{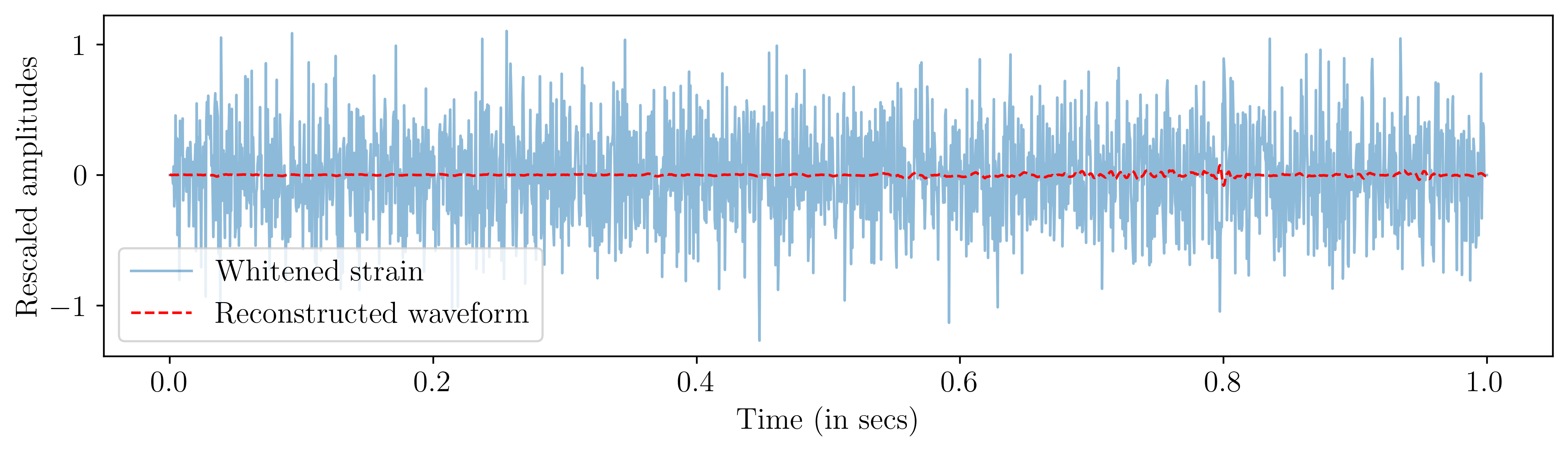}{0.8\textwidth}{(b)}}
\caption{(a) Reconstructed signal (red, dashed) corresponding to the injected waveform (green, solid) plotted against the whitened strain (blue, solid), obtained by injecting the simulated waveform in LIGO O3 noise. (b) Reconstruction of a pure noise sample with no GW injection.}
\label{fig:Example}
\end{figure}

Figure \ref{fig:Example} (a) and (b) show examples of an injection and a pure noise sample respectively, and their corresponding reconstructions obtained using \texttt{AWaRe}. The source parameters used to simulate the injection waveform are as follows: $m_{1} =$ 12.33 $M_{\odot}$, $m_{2}$ = 11.45 $M_{\odot}, |\vec{\chi}_{1}| = 0.75, |\vec{\chi}_{2}| = 0.48 $, spin polar angles, $\theta_{1} = 1.75 $ rad and $\theta_{2} = 2.05 $ rad, and spin azimuthal angles $\phi_{1} = 3.09 $ rad and $\phi_{2} = 0.47 $ rad. The SNR of the injection sample is 12. The overlap obtained for the injection samples is 0.91, indicating high degree of match between the reconstruction and the pure template. We notice that the amplitude and phase consistency between the reconstructed waveform and the template is better for oscillations around the merger where the signal is at its loudest, compared to the early inspiral and ringdown regions, which are comparatively much fainter. Figure \ref{fig:Example} (b) shows the whitened strain and \texttt{AWaRe}'s reconstruction for a pure noise sample. It was found that the model outputs a vector of small random fluctuations around zero amplitude for this sample, as is expected from non-injections. \\

\subsection{Injections in O3b noise}

%We have investigated \texttt{AWaRe}'s performance on a set of injections in O3b noise and plotted the overlaps as a function of the SNRs for different chirp masses, $\mathcal{M}_{c} = \frac{(m_{1}m_{2})^{3/5}}{(m_{1}+m_{2})^{1/5}}$ (Figure \ref{fig:Figure_3} (a)). We observe that for low SNRs, the overlaps are poor, indicative of the high uncertainty in detection and waveform reconstruction for faint signals (SNR $<$ 8). The overlaps improve with increasing SNR, with most of the SNR $\geq$ 12 injections having $\mathcal{O} $ $>$ 0.85. It is to be noted that the chirp masses of the injections with $\mathcal{O} $ $<$ 0.85 and SNR $\geq$ 12 mostly lie below 15 M$_{\odot}$. Since for smaller chirp masses, the lengths of the waveforms are longer, the model struggles to reconstruct the full waveform with high fidelity. In particular, the oscillations in the early inspiral stage are fainter compared to those around the merger, which causes the reconstructed waveforms to have mismatch in amplitiude and phase with actual templates. 

\begin{figure}
\centering
\includegraphics[scale=0.7]{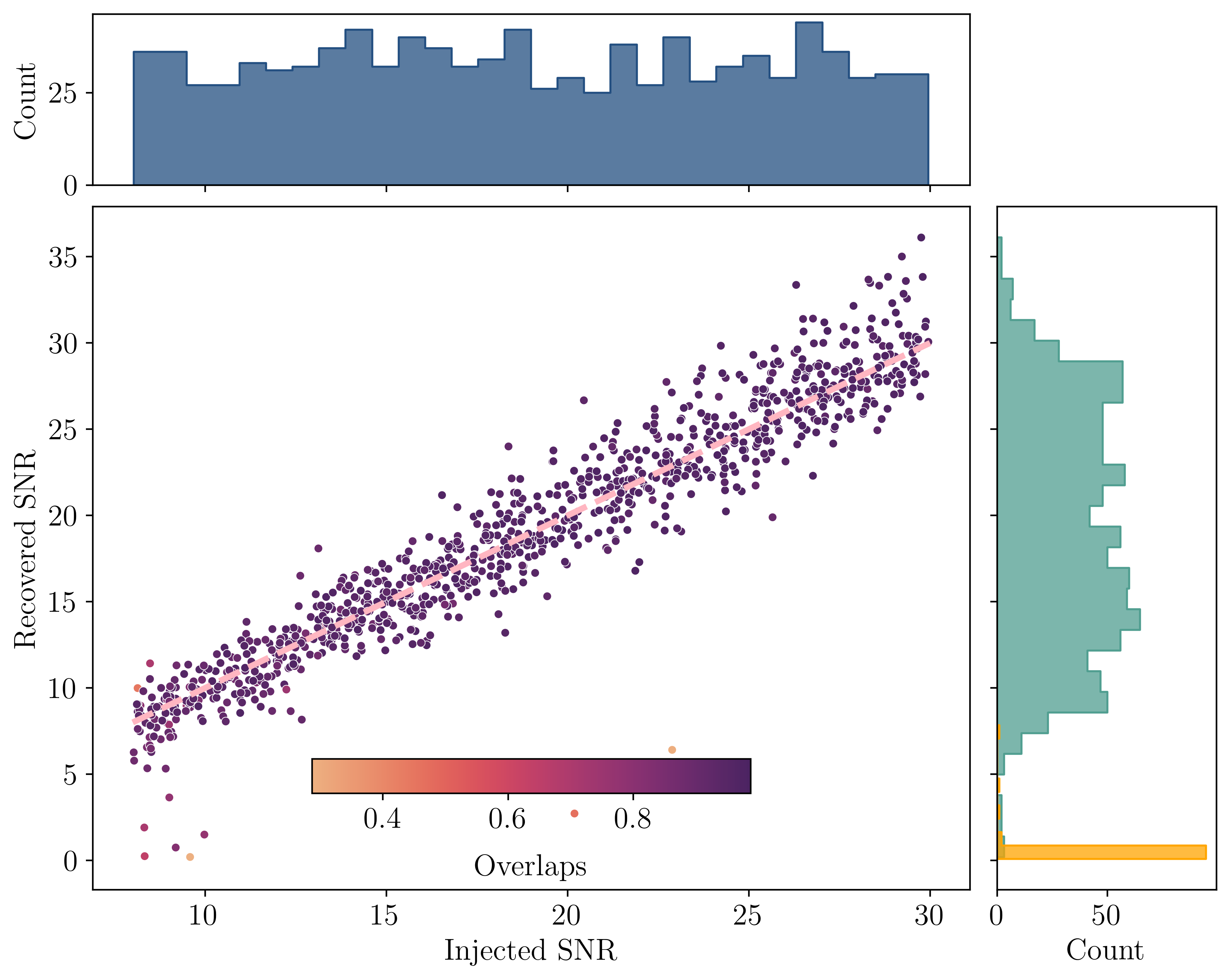}
%\gridline{\fig{Recovered_vs_Injected_SNR_snr>=8.png}{0.8\textwidth}{(a)}}
\caption{Recovered vs. injected SNR for test set injections. The coloured points indicate the overlaps. The blue straight line has been drawn for comparison. The solid histograms in the upper and right panels represent injected and recovered SNR counts respectively. The blue histogram on the right panel show recovered SNRs for pure noise samples, with no GW injections.}
\label{fig:Figure_3}
\end{figure}

We have investigated \texttt{AWaRe}'s performance on a set of injections in O3b noise having SNR between 8 and 30. Figure \ref{fig:Figure_3} illustrates the comparison between the injected and recovered SNRs for the test set, with the colours of the points indicating the overlaps between the reconstructed and true waveforms. We found that 86\% of the injections in our test set achieved more than 0.9 overlaps with the pure templates. For high overlap samples, the recovered SNR closely matches the injected SNR, with the overall distribution closely following the pink dashed curve along which the recovered and injected SNRs have the exact same values. The root-mean-square of residuals computed between the recovered and injected SNRs was found to be 1.98 and the variance was found to be 3.95. We observed that 0.7\% of the injections have both injected SNR $\geq $ 8 and recovered SNR $\leq $ 4, meaning in a real detection scenario, our model would have missed these events. Additionally, we compute the recovered SNRs for 100 pure noise samples (yellow histogram), which have no GW injections. The recovered SNRs for these samples are peaked at 0, which is what we expect from non-injection samples. \\

\begin{figure}
\gridline{\fig{Overlap_vs_Total_Mass}{0.50\textwidth}{(a)}
         {\fig{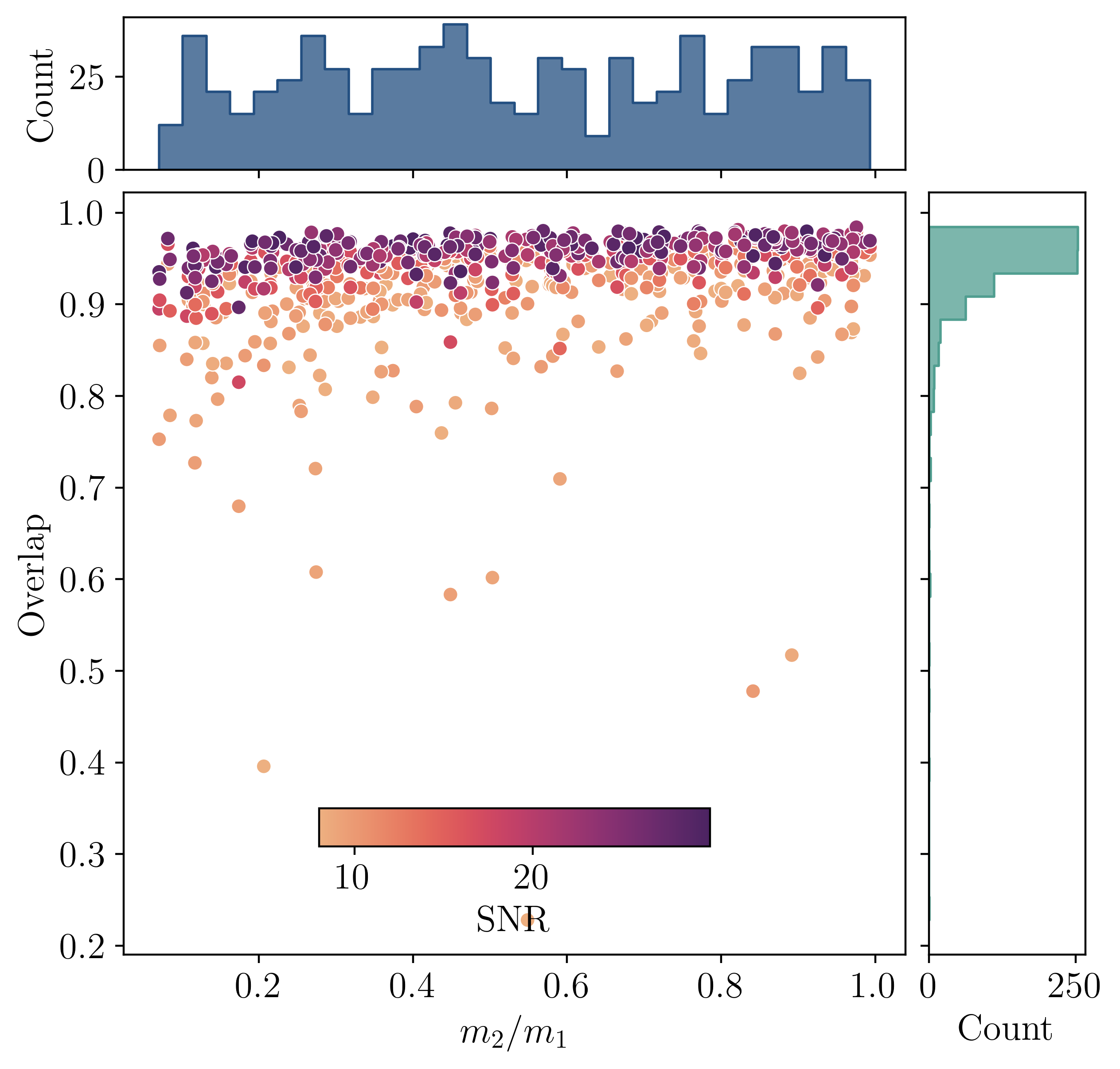}{0.50\textwidth}{(b)}}}
\gridline{\fig{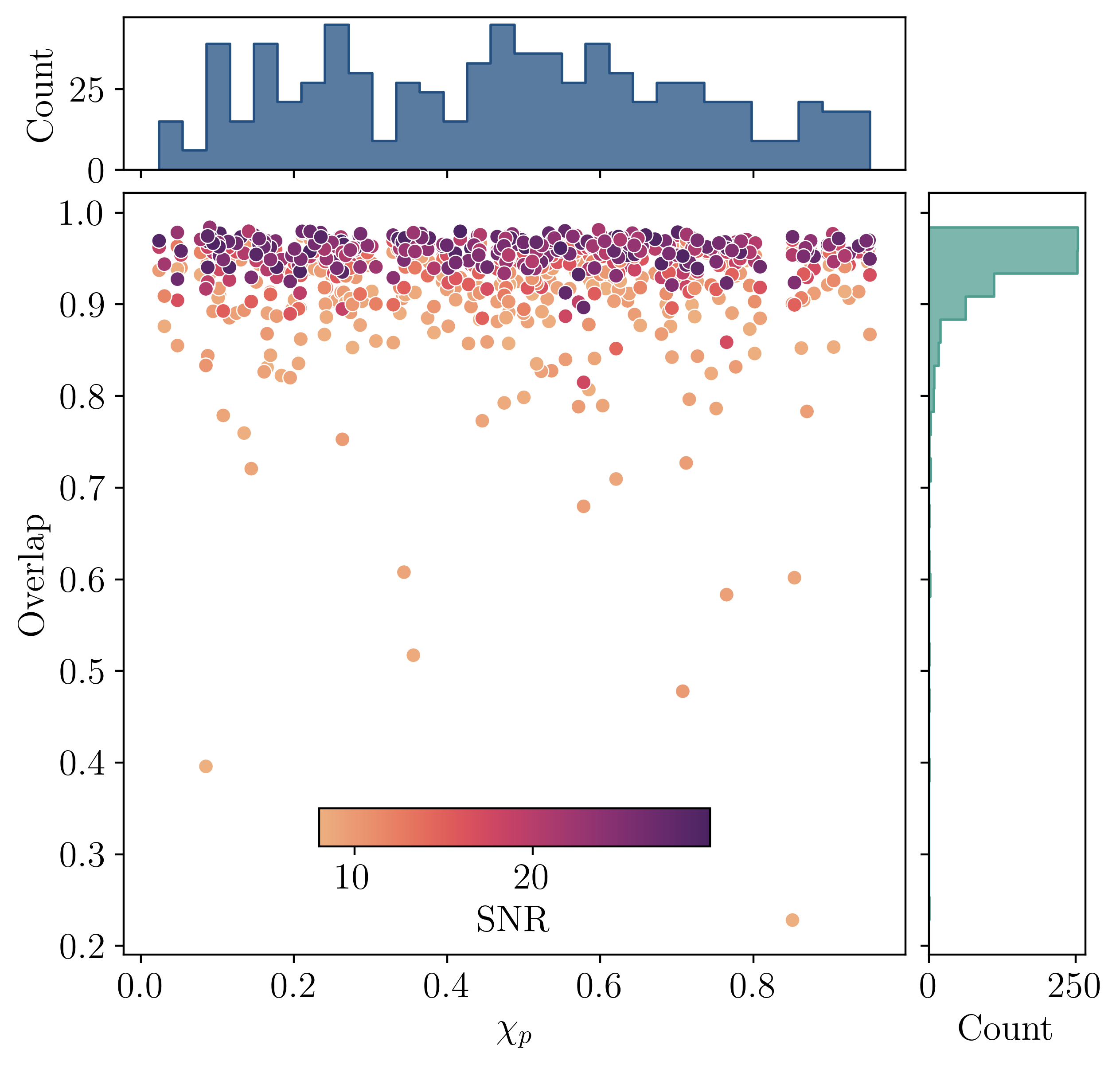}{0.5\textwidth}{(c)}
         {\fig{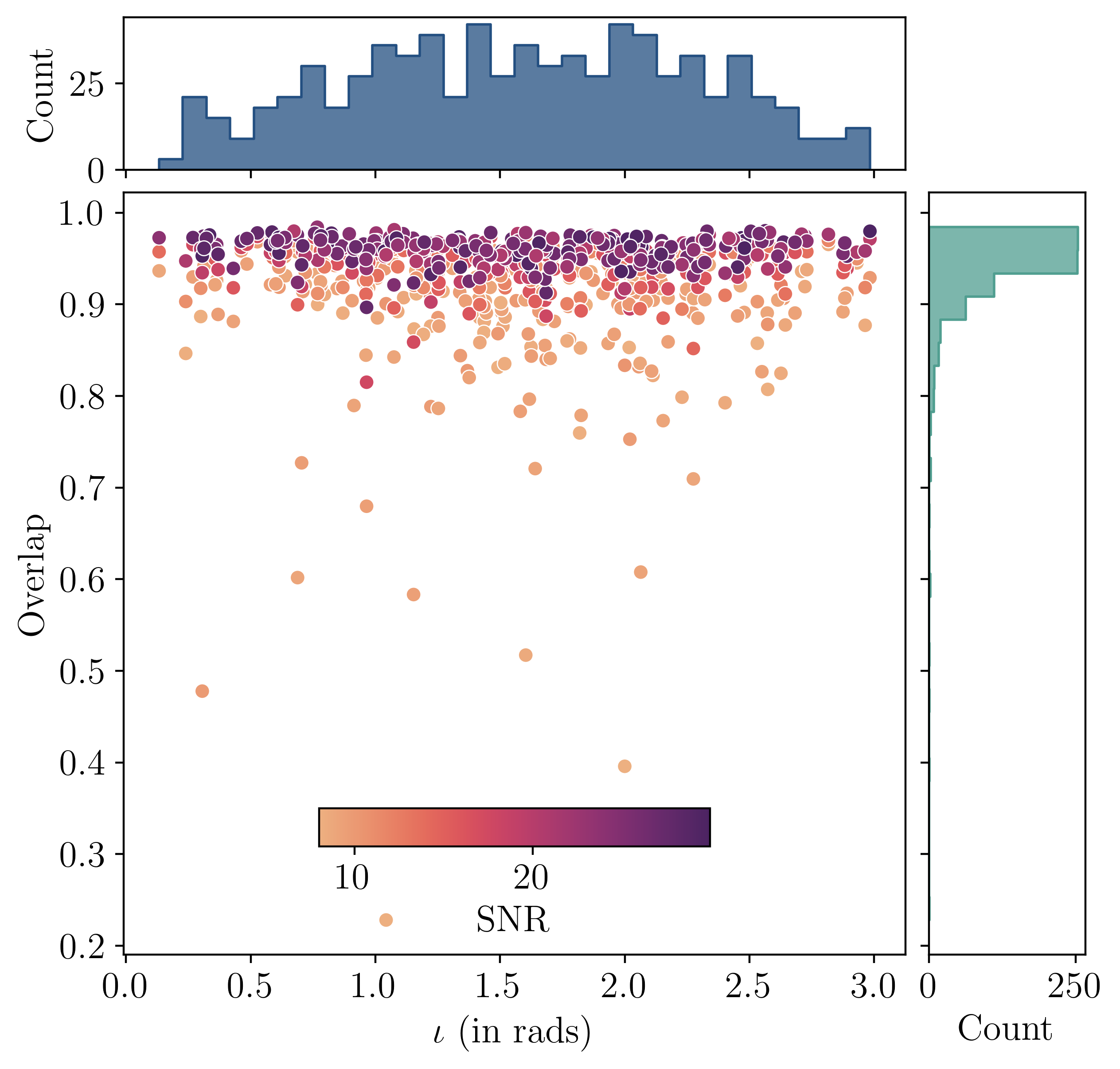}{0.50\textwidth}{(d)}}}
\caption{(a) Overlaps vs. total mass, (b) Overlaps vs. mass ratio, (c) Overlaps vs. effective spin precession and (d) Overlaps vs. inclination angle for a set of 1000 injection test samples. The colours of the points in all the figures represent the SNR, as shown in the colourbar. The histograms in the top panels show the counts of respective parameters and the ones on the right panels show the overlap counts.}
\label{fig:Overlaps_vs_HM_params}
\end{figure}

In Figure \ref{fig:Overlaps_vs_HM_params} (a) -- (d), we plot the overlaps against the total mass, $M$, the mass ratio, $m_{2}/m_{1}$, the effective spin precession parameter, $\chi_{p} $, and inclination angle, $\iota $ respectively, with the colours indicating the injection SNR. These parameters were chosen because they have a strong influence on the contribution of higher-order modes. Since higher-order modes only contribute significantly to the signal at the last few cycles, GWs from high total mass sources will have greater higher-order mode content because their frequency at merger and ringdown lie near the optimal sensitivity of the current detectors. Similarly, studies have shown that for unequal mass ratios, the $(l,m) = (3,3) $ and higher multipoles become important (\cite{Mass_ratio_HM_1, Mass_ratio_HM_2}). GWs from face-on binaries, with $\iota = 0$, are dominated by the (2,2) mode. But the effect of the higher multipoles increases as $\iota \to \pi/2$. A precessing orbital plane induces modulations in the frequency and amplitude of the $h_{l,m} $ modes computed in the non-precessing frame of reference. We notice across all the figures that for most of the injections with SNR $> $ 12, the overlaps are $>$ 0.9. This shows that the waveform reconstruction accuracy is largely unaffected by the  strength of the higher order mode contribution induced by the variation in parameters $M $, $m_{2}/m_{1}$, $\chi_{p} $, and $\iota $. \\
%Similarly, injections with smaller mass ratios tend to have lower overlaps, as shown in Figure \ref{fig:Figure_3} (b). This is likely because the contribution of the higher-order modes in the GW radiation becomes significant for more asymmetric mass binaries. Whereas for nearly equal masses, only the quadrupolar (2,2) modes dominate, the higher multipole contribution in asymmetric mass systems introduce notable alterations in both the amplitude and phase of the waveform. These complexities pose a greater challenge for the model's ability to accurately reconstruct these waveforms, which is manifested as a gradual decrease in overlap. \\

\begin{figure}
\centering
\includegraphics[scale=0.6]{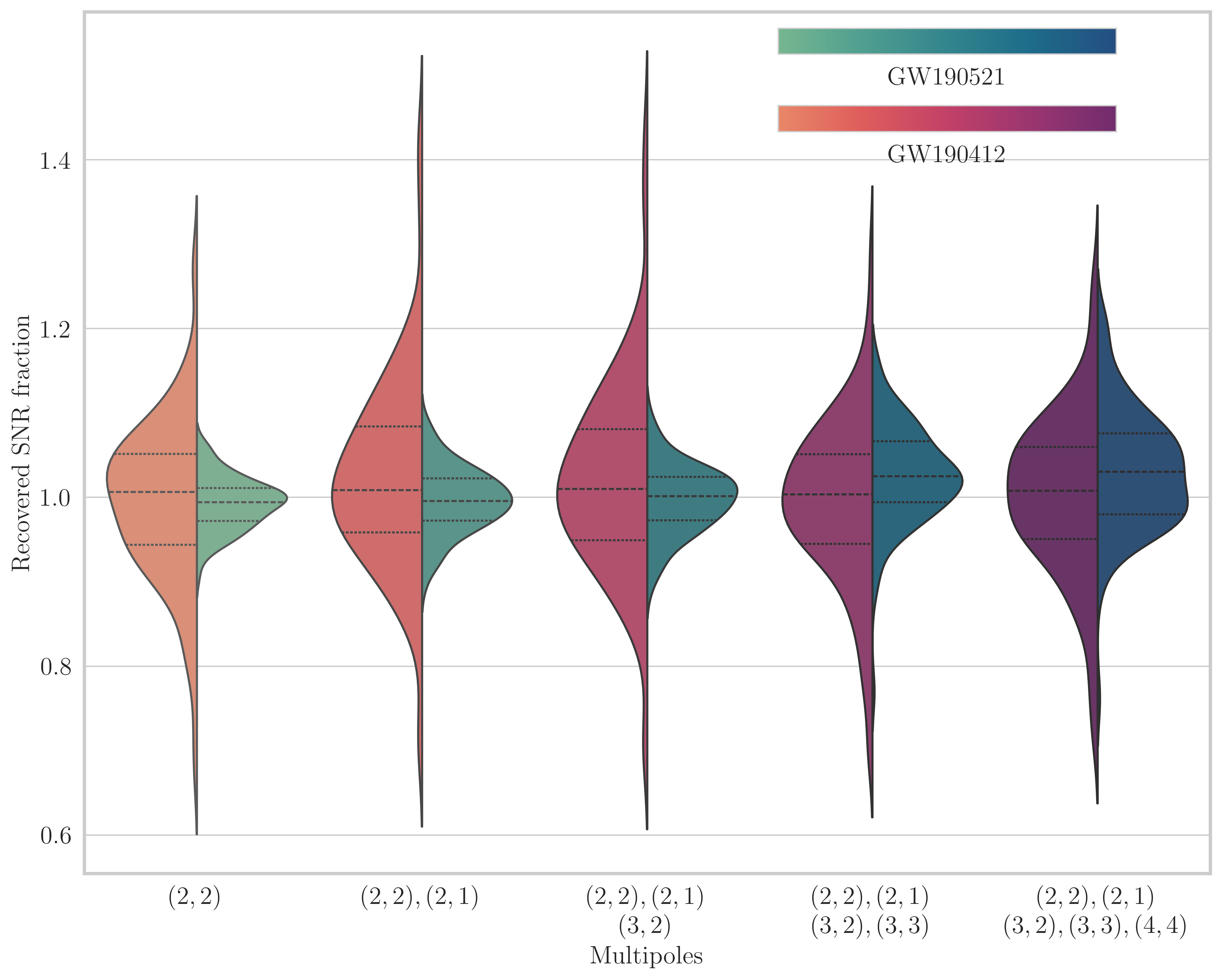}
%\gridline{\fig{GW190412_IMRPhenomXPHM_lm_modes_overlaps_violin.png}{0.5\textwidth}{(a)}
%         {\fig{GW190412_IMRPhenomXPHM_lm_modes_recovered_SNR_violin.png}{0.5\textwidth}{(b)}}
%         }
\caption{Violin plots showing distributions of the recovered SNR fraction (optimal SNR of reconstructed waveform/optimal SNR of original waveform) as a function of GW multipoles for 100 GW190521-like injections (green) and 100 GW190412-like injections (red). The dashed lines in the violin plots show the median and interquartile ranges.}
\label{fig:HM_injections}
\end{figure}

We have further tested the reconstruction accuracy of our model for incremental higher order mode content using simulated injections of GW190521 (\cite{GW190521}) and GW190412 (\cite{GW190412})-like events. Figure \ref{fig:HM_injections} presents violin plots showing distributions of recovered SNR fraction (defined here as the ratio of the optimal SNR obtained using the reconstructed waveform to the optimal SNR obtained using the pure templates) for different combinations of GW harmonics in GW190521 (green) and GW190412 (red). We chose GW190412 because the data for this event shows evidence for non-negligible higher order mode contribution, as reported in previous studies (\cite{GW190412}). The unequal masses ($q = m_{2}/m_{1} = 0.28^{+0.12}_{-0.07}$), non-zero $\chi_\text{{eff}} $ (= $0.25^{+0.08}_{-0.11}$) and $\chi_{p}  (=0.31^{+0.19}_{-0.16})$, make GW190412 the ideal test candidate for studying \texttt{AWaRe}'s higher order mode reconstruction accuracy. GW190521 is an interesting candidate for higher-order mode analysis because of its high total mass ($M = 150^{+29}_{-17} M_{\odot}$) and support for misaligned spins (Bayes factor = 11.5) \cite{GW190521}. Besides, a study by \cite{GW190521_HM} shows strong observational evidence for a multimode black hole ringdown spectrum for GW190521, with $(l,m) = (2,2)$ and $((l,m) = (3,3) $ being the dominant and sub-dominant modes respectively. A maximum Bayes factor of 56 $\pm $ (1$\sigma$ uncertainty) was obtained in this study, preferring two fundamental modes over one \cite{GW190521_HM}.\\

Each of the violin plots in Figure \ref{fig:HM_injections} show 100 injections simulated with the multipoles specified on the x-axis. We generated these waveforms with source parameters fixed at the median values of the posteriors of GW190521 and GW190412, and only varied the inclination angle by choosing a uniform prior over cos~$\iota$ between -1 and 1. We also set the distance of both events to 100 Mpc, in order to scale up the SNR, so that the higher order mode content becomes more prominent. The horizontal lines in the violin plots show the median and interquartile range. Since \texttt{AWaRe} has been trained on injections with all available higher-order modes in IMRPhenomXPHM, we find no significant difference in recovered SNR fraction between the datasets for the two events. For GW190521-like injections, we found a gradual increase in the interquartile range of the recovered SNR fraction around 1.0, starting from the inclusion of the (3,3) modes. This is likely because increasing higher-order mode content leads to complicated modulations in the amplitude and phase of the waveforms (see Fig.~1 in \cite{Bustillo_Nature}), making it challenging for the model to produce accurate reconstructions. Overall, the medians of the recovered SNR fractions of all the test sets lie close to 1.0, suggesting almost perfect SNR recovery. \\

%However, we notice a gradual increase in the interquartile range of the overlap distributions (represented by the height of the boxes in Figure \ref{fig:HM_injections} (a)) and an elongation of their minimum overlap tails after the addition of the (3,3) and (4,4) modes. This suggests that the (3,3) and higher multipoles have a significant effect on the morphology of GW190412-like events, making it increasingly challenging for \texttt{AWaRe} to produce reconstructions that have similar accuracy to the $h_{22}$ waveforms. 

%\begin{figure}
%    \centering
%    \includegraphics[scale=0.60]{Recovered_vs_Injected_SNR.png}
%    \caption{Recovered vs. injected SNR for test set injections. The coloured points indicate the overlaps. The blue straight line have been drawn for comparison. The solid histograms in the upper and right panels represent injected and recovered SNR counts respectively. The blue histogram on the right panel show recovered SNRs for pure noise samples, with no GW injections.}
%    \label{fig:SNR_comparison}
%\end{figure}

%\begin{figure}
%\gridline{\fig{GW190521_IMRPhenomXPHM_lm_modes_overlaps.png}{0.5\textwidth}{(a)}
%         {\fig{GW190412_IMRPhenomXPHM_lm_modes_overlaps.png}{0.5\textwidth}{(b)}}
%         }
%\gridline{\fig{GW190521_IMRPhenomXPHM_lm_modes_recovered_SNR.png}{0.5\textwidth}{(c)}
%         {\fig{GW190412_IMRPhenomXPHM_lm_modes_recovered_SNR.png}{0.5\textwidth}{(d)}}
%         }
%\caption{(a) Overlaps between reconstructed waveforms and pure templates as a function of SNR for 2000 injection samples. The colours of the points indicate the chirp mass. (b) Overlaps as a function of SNR for different mass ratios, indicated by the coloured points. The panels on the right of (a) and (b) show the histograms of the overlap values.}
%\label{fig:HM_injections}
%\end{figure}

\subsection{Real GW events}

\begin{figure}
\gridline{\fig{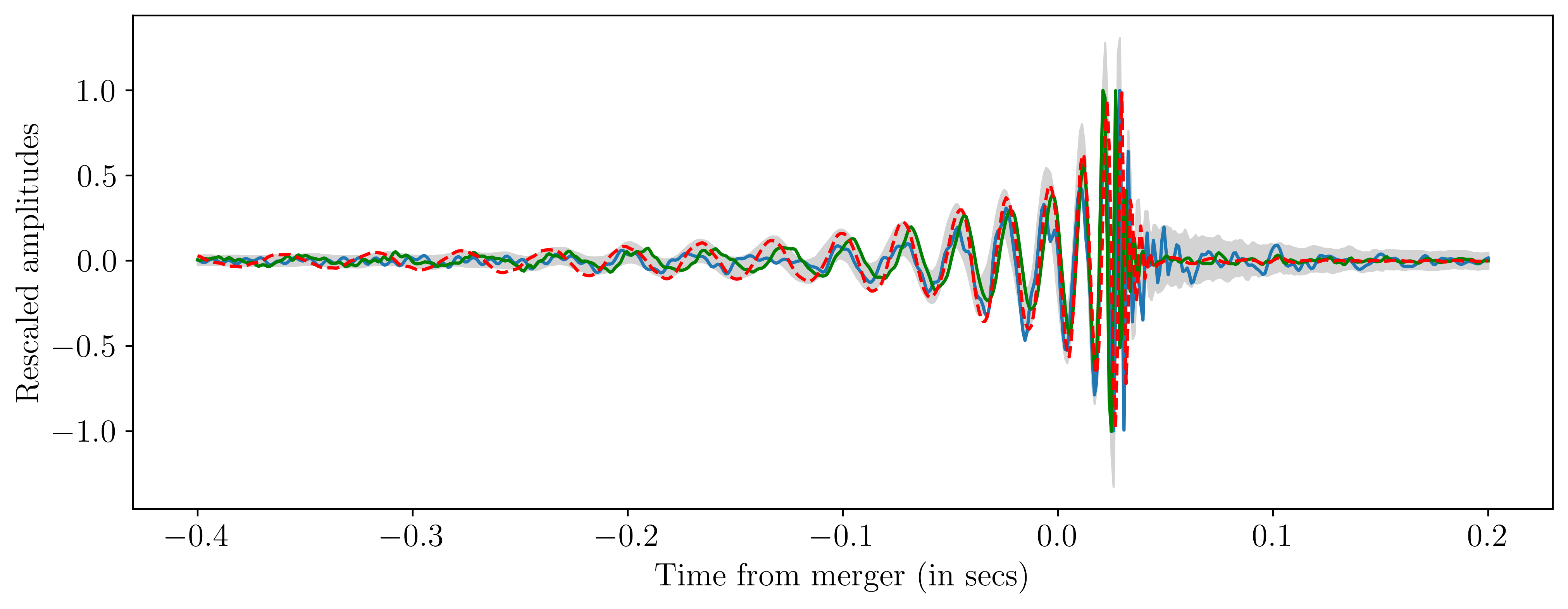}{0.5\textwidth}{(a)}
         {\fig{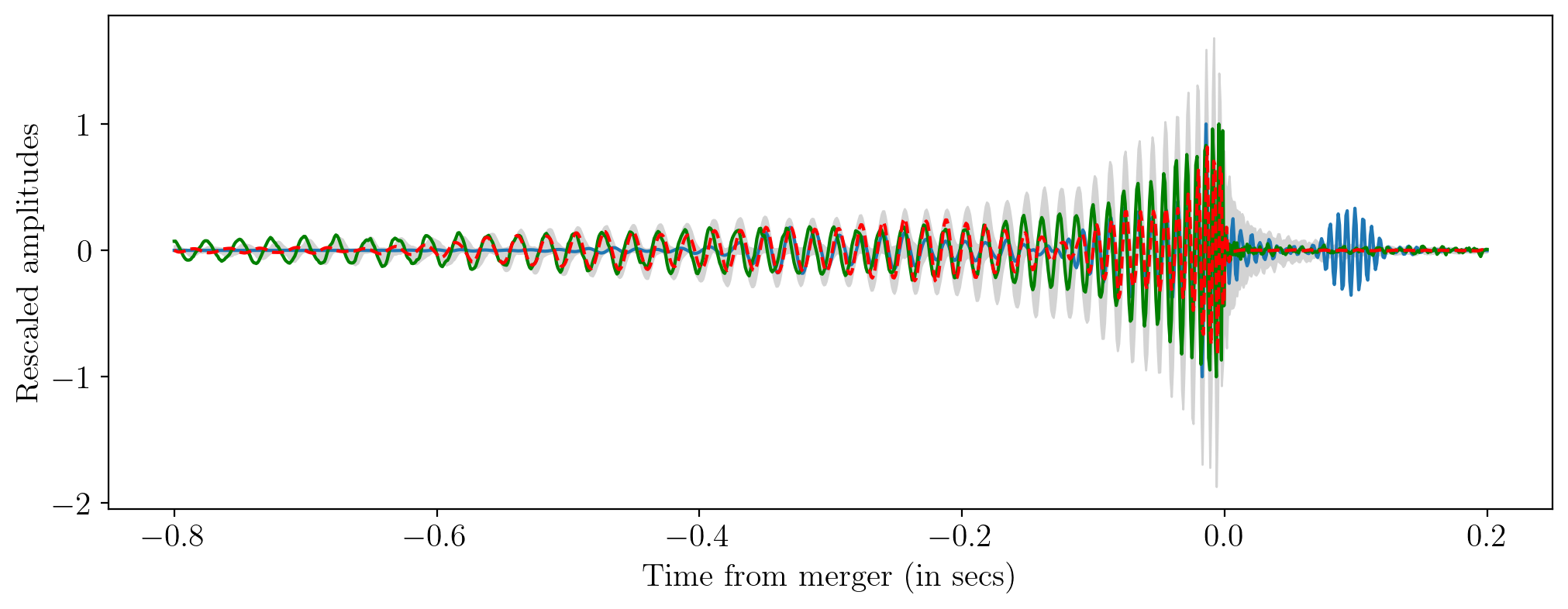}{0.5\textwidth}{(b)}}}
\gridline{\fig{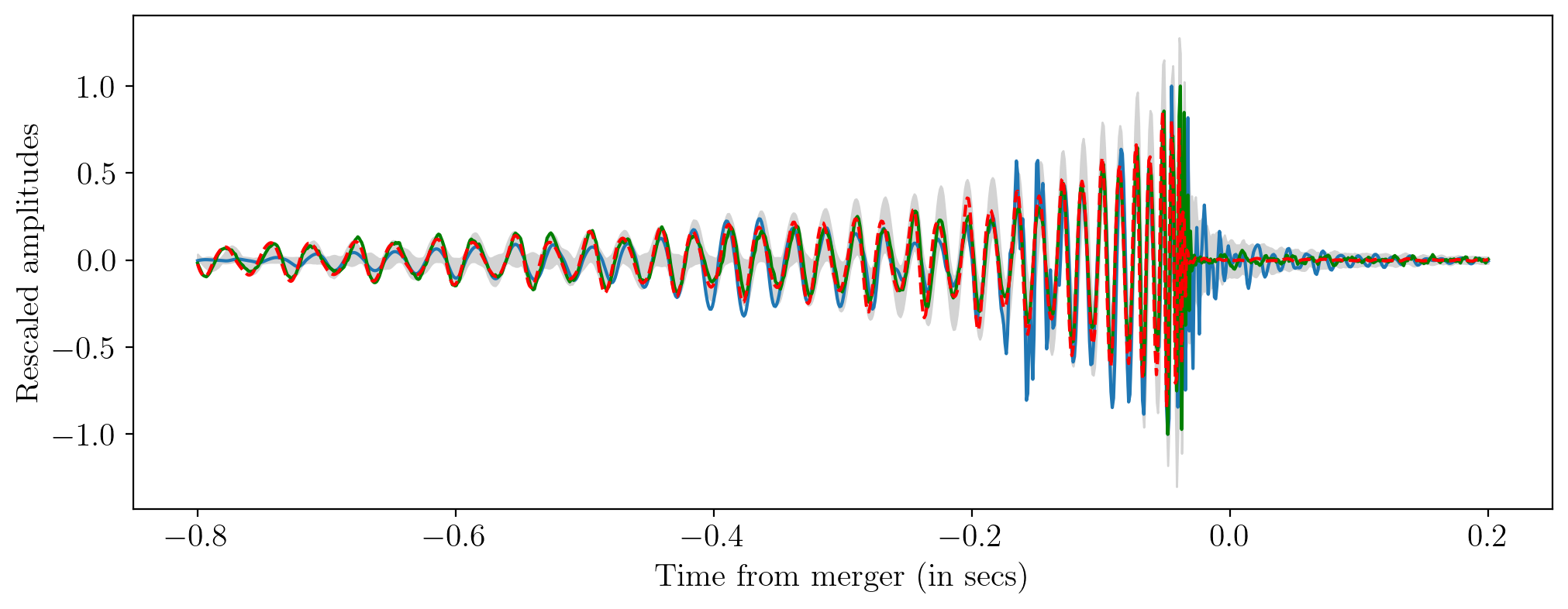}          {0.5\textwidth}{(c)}
         {\fig{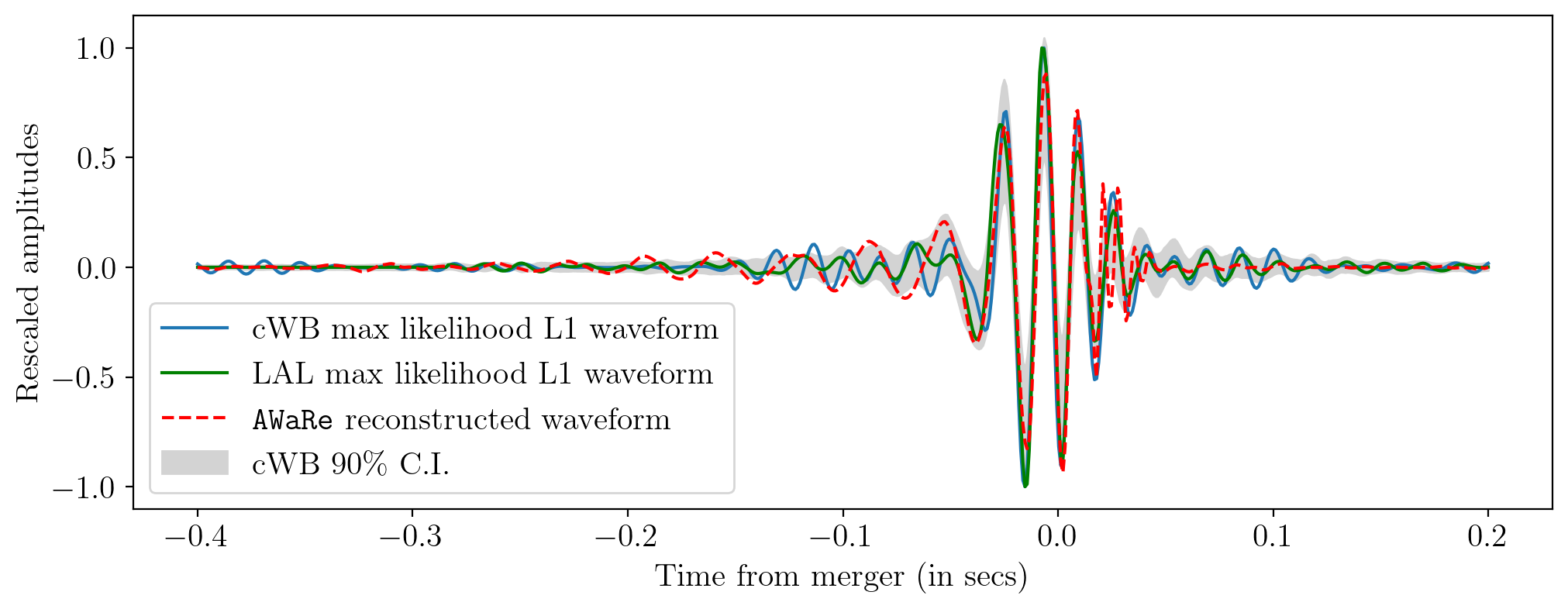}{0.5\textwidth}{(d)}}}
\caption{Clockwise from top left: Comparison of Livingston signal reconstructions by the \texttt{AWaRe} model, cWB and LALInference for GW150914, GW151226, GW190521 and GW190412. The \texttt{AWaRe} predictions are shown in red, dashed lines. The LALInference and cWB reconstructions are shown in green and blue respectively. The gray regions show the cWB 90\% confidence intervals.}
\label{fig:Real_events}
\end{figure}

We have tested the performance of \texttt{AWaRe} on interesting GW events detected during the first three observation runs of LIGO and Virgo, and compared our model's reconstructions with the maximum likelihood waveforms obtained by cWB and LALInference. Figures \ref{fig:Real_events} (a) -- (d) show results for the Livingston signal reconstructions of the events GW150914 (\cite{GW150914}), GW151226 (\cite{GW151226}), GW190412 (\cite{GW190412}) and GW190521 (\cite{GW190521}) respectively. The \texttt{AWaRe} reconstructions are shown using red, dashed lines, while the green and blue lines show the maximum likelihood reconstructions of LALInference and cWB. The data of the cWB and LALInference reconstructions and 90\% confidence intervals were obtained from \cite{cWB_results}. \\

%The gray shaded areas show the cWB 90\% confidence intervals  obtained by adding waveforms to data near, but not including the event. The cWB analysis on these `off-source' injections was repeated thousands of times to estimate the distributions of the reconstruction parameters. \\

We found that for all the real events considered in the analysis, the reconstructions obtained by our model are within the cWB 90\% confidence intervals and have excellent match with the maximum likelihood waveforms from both cWB and LALInference. The overlaps and the recovered SNR fraction relative to cWB and LALInference are shown in Table \ref{tab:Real_events}. GW150914, being the first GW signal ever detected by LIGO, and one of the loudest events detected to this day, yields very high overlaps and recovered SNRs close to unity when tested with \texttt{AWaRe} (Figure \ref{fig:Real_events} (a)). Similar performance is expected for most vanilla BBHs within similar SNR ranges detected in the last three observation runs. Since cWB is designed for unmodelled burst searches, its reconstruction of low chirp mass sources, having $> $ 1 sec long waveforms, like GW151226  and GW190412, are poorer compared to LALInference and \texttt{AWaRe}, which have much better match (Figure \ref{fig:Real_events} (b) and (c)). For GW190521, which has total mass $> $ 100 M$_{\odot}$, the cWB reconstructions show larger number of cycles both prior and post-merger, compared to LALInference (Figure \ref{fig:Real_events} (d)). This is owing to the model-agnostic nature of the cWB algorithm, which is better at capturing the amplitude and phase modulations induced by higher-order modes, expected to have significant impact in the evolution of the waveform in these high-mass systems. For all the real events we have tested on, we find a mismatch in amplitude and phase between the reconstructed ringdown oscillations with the cWB and LALInference results. This is owing to the very low amplitudes of the waveform around this region, making it difficult for the model to extract the waveform with high fidelity. For the LALInference analysis, the waveform model used for all the events was IMRPhenomPv2 (\cite{IMRPhenomPv2}), except for GW190521 and GW190412, for which IMRPhenomPv3HM (\cite{IMRPhenomPv3HM}) was used.

\begin{table}[h]
\caption{Comparison of reconstructed waveforms from \texttt{AWaRe}, cWB and LALInference for real GW events detected during O1-O3. The SNR fraction refers to the ratio of the optimal SNR from the reconstructed waveforms to that of the pure templates.}
\begin{tabular}{lllll}
\hline
\textbf{Event ID} & \textbf{Overlap (cWB)} & \textbf{Overlap (LAL)} & \textbf{SNR fraction (cWB)} & \textbf{SNR fraction (LAL)}\\ 
\hline
GW150914 & 0.92 & 0.98 & 1.1 & 1.15 \\
GW151226 & 0.79 & 0.84 & 1.14 & 0.7\\
%GW170729 & 0.84 & 0.89 & 0.96 & 1.04\\
GW190412 & 0.74 & 0.91 & 0.9 & 1.03\\
GW190521 & 0.86 & 0.89 & 0.94 & 0.98\\
%GW190602\_175927 & 0.96 & 0.96 & 0.96 & 0.99\\
\hline
\end{tabular}
\label{tab:Real_events}
\end{table}

%GW190521: Overlap (cWB) = 0.86, Overlap (LAL) = 0.89, SNRfrac (cWB) = 0.94, SNRfrac (LAL) = 0.98 \\
%W190602: Overlap (cWB) = 0.96, Overlap (LAL) = 0.96, SNRfrac (cWB) = 0.96, SNRfrac (LAL) = 0.99 \\
%GW190412: Overlap (cWB) = 0.74, Overlap (LAL) = 0.91, SNRfrac (cWB) = 0.9, SNRfrac (LAL) = 1.03 \\
%GW170729: Overlap (cWB) = 0.84, Overlap (LAL) = 0.89, SNRfrac (cWB) = 0.96, SNRfrac (LAL) = 1.04 \\
%GW151226: Overlap (cWB) = 0.79, Overlap (LAL) = 0.84, SNRfrac (cWB) = 1.14, SNRfrac (LAL) = 0.7 \\

\section{Conclusion}

We have developed a deep learning model for waveform reconstruction of Binary Black Hole (BBH) Gravitational Wave (GW) signals, which include higher-order modes, emitted from the merger of precessing binaries. Given a noisy strain from the LIGO-Virgo interferometers, this model outputs the whitened waveform, effectively removing the noise and revealing the underlying real waveform. For inputs consisting of pure noise without any GW signal, the model outputs a vector of random fluctuations around zero, indicating the absence of true astrophysical signals. This method utilizes an attention module to enhance feature learning from data, an idea inspired by state-of-the-art natural language processing models. \\

We have demonstrated the feasibility of this approach on injections in actual LIGO O3 noise, as well as on real GW data, and compared our results with those from cWB and LALInference. Our systematic study on the reconstruction accuracy of our model against a range of parameters influencing higher-order modes shows robust performance in systems with SNRs greater than 15. Thus, this model can serve as a rapid trigger generation algorithm to flag data segments potentially containing GW signals, which can then be further analyzed using matched filtering. This application will not only save the computationally intensive process of performing continuous matched filtering on live detector data but will also aid in detecting signals with higher-order modes and precession, which are currently not included in the online GW search pipelines. While the unmodeled search pipeline, cWB, can detect such exotic GW sources, it does not provide robust real-time estimates of signal reconstruction uncertainty. Certain modifications to our deep learning approach, such as the use of variational autoencoders or Bayesian neural networks, can address this issue by providing both a point estimate and corresponding uncertainty for the reconstructed waveform prediction. We aim to explore this in our future work.

\begin{acknowledgments}
This research was undertaken with the support of compute grant and resources, particularly the DGX A100 AI Computing Server, offered by the Vanderbilt Data Science Institute (DSI) located at Vanderbilt University, USA. This research used data obtained from the Gravitational Wave Open Science Center (https://www.gw-openscience.org), a service of LIGO Laboratory, the LIGO Scientific Collaboration and the Virgo Collaboration. LIGO is funded by the U.S. National Science Foundation. Virgo is funded by the French Centre National de Recherche Scientifique (CNRS), the Italian Istituto Nazionale della Fisica Nucleare (INFN) and the Dutch Nikhef, with contributions by Polish and Hungarian institutes. This material is based upon work supported by NSF's LIGO Laboratory which is a major facility fully funded by the National Science Foundation. 
\end{acknowledgments}

\bibliography{sample631}{}
\bibliographystyle{aasjournal}



\end{document}